\documentclass[10pt,letterpaper,twocolumn]{article} 

\usepackage{ol2}
\usepackage[draft]{hyperref}
\usepackage{amsmath}

\begin{document}

\twocolumn[ 

\title{Efficient output coupling of intracavity high harmonic
generation}


\author{D. C. Yost$^\dagger$, T. R. Schibli, and Jun Ye}

\address{
JILA, National Institute of Standards and Technology and University
of Colorado\\Department of Physics, University of Colorado, Boulder,
Colorado 80309-0440, USA}

\begin{abstract}

We demonstrate a novel technique for coupling XUV harmonic radiation
out of a femtosecond enhancement cavity.  We use a small-period
diffraction grating etched directly into the surface of a dielectric
mirror. For the fundamental light, this element acts as a high
reflector.  For harmonic wavelengths, it acts as a diffraction
grating, coupling XUV radiation out of the cavity. Using this
method, we observed the $3^{\rm rd}$ through $21^{st}$ odd harmonics
with a dramatic increase in usable power over previous results of
high harmonic generation at high repetition rates.

\end{abstract}


] 
\noindent
Recent developments in optical frequency combs have revolutionized
optical frequency metrology \cite{Udem02,Cundiff03}. However these
techniques have in general been limited to the visible and near IR
spectral regions. Meanwhile, progress in short-wavelength light
sources has been rapid, achieving unprecedented temporal resolution,
spectral coverage, and brightness \cite{Brabec}. A powerful technique 
for producing XUV wavelengths is high harmonic generation (HHG) which 
utilizes extreme nonlinear optical processes in atoms and molecules 
facilitated by amplified femtosecond pulses \cite{Corkum, Lewenstein,Spielmann}.
Traditional methods for generating pulses of
sufficient energy for HHG employ low repetition rate amplifiers,
leaving no comb structure in the spectrum of the harmonic radiation.
Therefore, the spectral resolution of these short-wavelength sources 
is poor when compared with precision visible sources.

Experiments in which broadband, femtosecond pulses are coupled into
passive external cavities are very promising in this respect as the
pulse energy is sufficiently enhanced to enable the HHG process
without a decrease in the pulse repetition rate \cite{Jones2}.
Already, there have been two successful efforts in intracavity high
harmonic generation with explicit demonstrations of phase coherence
of the $3^{rd}$ harmonic light, resulting in high expectations that
this method can successfully push frequency comb techniques into the
XUV spectral region \cite{Jones1, Gohle}. Especially intriguing is
the prospect of applying the recently established technique of
direct frequency comb spectroscopy (DFCS) \cite{Marian} in the XUV
spectral region \cite{Bellini, Witte2}.

One of the outstanding technical challenges to generating XUV
radiation via an enhancement cavity has been the lack of a suitable
method to couple the harmonic light out of the cavity.  Inside the
enhancement cavity, the harmonics are generated collinearly with the
fundamental light. Since there is essentially no solid material that
is sufficiently transparent at XUV wavelengths, the XUV radiation
can not pass through the cavity mirrors without being absorbed. 
A requirement for any output coupling method is that its
implementation should not significantly increase the cavity loss nor
should it introduce significant nonlinearity that prevents
efficient couplings between a train of ultrashort pulses and the
cavity.  Thus far,
proposed XUV output coupling methods have been less than desirable
in this regard. Published results have shown output coupling of
harmonic radiation using a thin sapphire plate at Brewster's angle
\cite{Jones1, Gohle}. Since the index of the sapphire is less than
unity at the harmonic wavelengths, one can achieve a large Fresnel
reflection of the harmonic radiation while introducing only a small round
trip loss to the fundamental pulses. This method has a fundamental
limitation, however, in that $\chi^{(3)}$ nonlinear processes within the
Brewster's plate introduce prohibitively large dispersion at high
intensities, severely limiting the possibility of power scaling such a
system \cite{Moll1}.

Another proposed method for coupling harmonics out of the
enhancement cavity is to drill a small hole in the curved mirror
after the intracavity focus \cite{Moll2}.  Since the harmonic light
will diverge less than the fundamental light, most of the harmonic
light will pass through the hole while most of the fundamental light
will be reflected.  While this method allows for power scaling, the
small hole invariably introduces additional loss to the cavity. This
intracavity loss can be somewhat mitigated by coupling a
higher-order transverse mode (e.g. TEM$_{01}$) into the cavity at
the cost of losses outside the cavity due to mode conversion.  Even
so, in practice we have found it difficult to keep intracavity
losses small using this method. Also, the harmonic radiation generated
with a TEM$_{01}$ fundamental beam will display a complicated
transverse mode profile that is not ideal for subsequent
experiments. A similar proposed technique uses a slotted mirror and
two colliding pulses to produce non-collinear high harmonic
generation \cite{Moll2, Wu}. This method also suffers from
additional cavity loss for the fundamental light and the implementation 
is very challenging technically.

\small
\begin{figure}[htb]
\centerline{
\includegraphics[width=8.3cm]{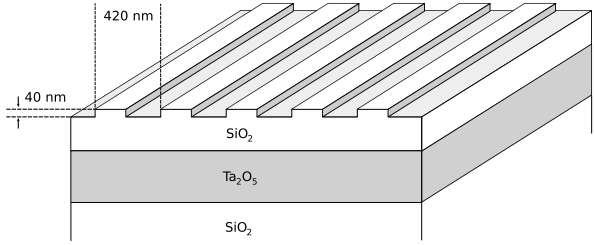}}
 \caption{Cross section of the grating used to couple harmonic radiation out of
the enhancement cavity.  The grating is etched directly into the surface of a
dielectric coating designed for high reflectivity and low dispersion at
1070~nm. The reflectivity of the dielectric mirror is 99.988\%.}
\end{figure}
\normalsize

In this Letter we demonstrate a novel method for coupling harmonics
out of an enhancement cavity that is far superior to the Brewster's
plate method in terms of power scalability. It also introduces
substantially smaller losses than coupling the harmonic radiation out of
the cavity through a small hole in a curved mirror. We add an optic
directly after the intracavity focus. The element acts as a
diffraction grating for all harmonic orders and yet, is a near
perfect high-reflector for the 1070~nm fundamental light.  We find
this optic produces no measurable decrease in the cavity finesse and
power enhancement. It also seems to have a minimal effect on the
cavity dispersion as determined from the measured transmitted
spectrum of the fundamental light. To manufacture this crucial
element, we began with a dielectric mirror coating, engineered for
low dispersion and high reflectivity at 1070~nm and a $70^\circ$
angle of incidence with S-polarization. In the top coating layer we
had a 420~nm period diffraction grating etched into the fused silica
surface. In Fig. 1 we depict the cross section of this optic.  The
period is small enough so that the fundamental 1070~nm light has
only a zeroth diffracted order. While this introduces birefringence
in the top layer of the coating \cite{Haggans}, the operation of the
mirror at the fundamental wavelength is left largely unaffected. The
high harmonics have higher diffracted orders with the largest power
being diffracted into the negative first order.

The operation of this optic at XUV wavelengths relies on a simple Fresnel reflection from the surface
of the grating structure as there is strong absorption in the bulk
of the material.  Even so, Fresnel reflections
from a vacuum/SiO$_2$ interface can be very large at XUV
wavelengths with a grazing angle of incidence.  In fact, at the
$9^{th}$ through $19^{th}$ harmonics, the Fresnel reflection is
$\sim45\%$ with S-polarization at a $70^\circ$ angle of incidence.
It might be interesting to increase this angle further to help
improve the output coupling efficiency. However, cavity geometries
under this circumstance will be more difficult to implement.

To estimate the output coupling efficiency of the intracavity
grating, we used an approximation of the rigorous integral method
described in \cite{Goray} with the index and absorption values taken
from \cite{Palik}.  We maximized the efficiency for the $9^{th}$
through $19^{th}$ harmonics by adjusting the period, duty cycle and feature
height 
of the grating with the requirement that the fundamental light diffracts only
into the zeroth order.  
The final grating design had a
period of 420~nm, a duty cycle of 40\%, and a step height of 40~nm
as shown in Fig. 1. The calculated output coupling efficiencies are
shown in Table 1.  Due to limitations in the manufacturing
capability, the grating structure we used had a 47\% duty cycle
which affected the performance minimally. The grating output coupling
efficiencies are very competitive with the other proposed and
demonstrated methods without the drawbacks mentioned previously.  
To increase the efficiency for a specific harmonic order using this method, 
one could attempt to etch an appropriate blazed structure into the surface of a high reflector.

\small
\begin{figure}[htb]
\centerline{
\includegraphics[width=8.3cm]{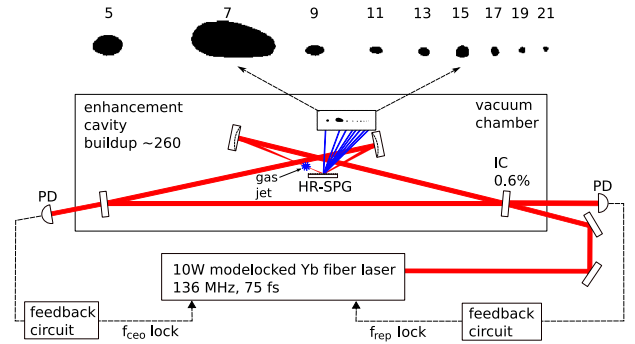}}
\caption{Experimental setup: IC, input coupler; PD, photodiode;
HR-SPG, high-reflector with small period grating on surface. The top
panel shows an enlarged image of the experimentally observed high
harmonic orders fluorescing on a glass plate coated with sodium
salicylate. The threshold of the image was reduced around the
$21^{st}$ harmonic to show its mode structure.}
\end{figure}
\normalsize

To experimentally evaluate the performance of this output coupling
method, we build on previous work where we resonantly enhanced light
from an amplified mode-locked Yb-fiber laser in a passive
high-finesse cavity \cite{Hartl}.  The reconfigured cavity
incorporating the grating is shown in Fig. 2. Before enhancement,
the laser produces 10~W of power with $\sim$75~fs pulses at a
136~MHz repetition rate. To coherently add the femtosecond
pulses inside the cavity, we control both independent degrees of
freedom of the frequency comb, the repetition rate $f_{rep}$ and the
offset frequency $f_{ceo}$ \cite{Jones2}.  On resonance, we achieve
an enhancement of 260 to obtain 2.6~kW of intracavity power and an
intracavity pulse duration of $\sim100$~fs. The focus in our cavity
is achieved with two 10-cm radius of curvature mirrors which produce a calculated
focused spot area of $960~\mu \rm{m}^2$ and hence a peak intensity of $4
\times 10^{13}$~W/cm$^2$.  We inject xenon gas at the intracavity
focus using a glass nozzle with an aperture of $100~\mu$m and a
backing pressure of 700-1500~Torr. The intensity is large enough to
produce high harmonic radiation in the xenon gas, which is subsequently
coupled out of the cavity by the grating.  The power of the
harmonics is measured with a calibrated photodiode \cite{IRD} and also observed
visually with a fluorescent plate.

We were able to simultaneously observe all odd harmonic orders
up to the $21^{st}$ through fluorescence on the coated glass plate.
Photodiodes sensitive at XUV wavelengths with directly deposited
filters were used to block background fundamental light by many
orders of magnitude, allowing measurement of the harmonic power.
Unfortunately, we had no metallic filters available which
efficiently transmit wavelengths longer than $\sim100$~nm.  This
severely limited our ability to measure the power in the harmonics
below the $13^{th}$.  We were able to measure the power in the
$13^{th}$ harmonic with an XUV-sensitive diode coated with 200~nm of
In and 20~nm of MgF$_2$.  The power in the $15^{th}$ through
$19^{th}$ harmonics was measured using a photodiode with a 150~nm
aluminum coating. The results of these power measurements are shown
in Table 1.  The cutoff of our harmonic radiation was observed to lie between
the $19^{th}$ the $21^{st}$ harmonics.  The power level in the $13^{th}$ harmonic shows an intracavity conversion efficiency of $\sim10^{-9}$ (corresponding to $\sim3\times10^{-7}$ for the unenhanced power).  The observed conversion efficiency, cutoff
wavelength and relative power levels of the harmonic orders fit well with theoretical calculations at our intensity levels \cite{L'Huillier}.

\begin{table}
  \centering
  \caption{Theoretical efficiency of the grating at harmonic wavelengths of the
incident 1070~nm light and the respective out-coupled
powers measured in the experiment.}\begin{tabular}{ccccc} \\ \hline
    harmonic & $\lambda(nm)$ & coupling & out-coupled &  \\
    order & & efficiency & power (nW)&  \\ \hline
    3 & 356.7 & 0.7\% & - &  \\
    5 & 214.0 & 1.9\% & - &  \\
    7 & 152.9 & 3.5\% & - &  \\
    9 & 118.9 & 9.6\% & - &  \\
    11 & 97.3 & 8.5\% & - &  \\
    13 & 82.3 & 9.0\% & 250 &  \\
    15 & 71.3 & 10.3\% & 20 &  \\
    17 & 62.9 & 9.6\% & 54 &  \\
    19 & 56.3 & 8.4\% & 38 &  \\
    21 & 51.0 & 6.7\% & - & \\ \hline
  \end{tabular}
\end{table}

To our knowledge, this is the first demonstration of the $17^{th}$
through $21^{st}$ harmonic orders via intracavity HHG.  It is notable to compare the measured power in
the $17^{th}$ harmonic at 63~nm with previous results at 61~nm (the
$13^{th}$ harmonic of 795~nm fundamental light), which displays an
increase in the output coupled power by nearly four orders of
magnitude\cite{Gohle, Jones1}.  We
attribute this dramatic increase in usable harmonic power to the
ability of using a laser with higher peak powers for the intracavity
experiment without being plagued by nonlinear dispersions introduced
by a Brewster's plate.

To conclude, we have demonstrated a novel method for output coupling
harmonics from a femtosecond enhancement cavity and in the process
have shown record power levels and the highest harmonic orders ever
produced at multi-megahertz repetition rates.  The
small-period intracavity grating has overcome the major difficulties
of coupling high harmonic radiation out of an optical buildup cavity. It
adds minimal intracavity loss, permitting a large buildup peak power
without introducing nonlinear phase shifts to the cavity resonance.
The absence of nonlinearities within the enhancement cavity removes
one of the major barriers to power scaling such a system.

We gratefully thank I. Hartl, A. Marcinkevi\v{c}ius and M. Fermann at IMRA America, Inc. for the design and construction of the
high-power Yb-fiber laser system. Funding is provided by AFOSR, DARPA, NIST and NSF.

\noindent
$^\dagger$dylan.yost@colorado.edu


\end{document}